\documentclass[12pt]{amsart}

\usepackage{graphicx}
\usepackage{amsmath}
\usepackage{amssymb}
\usepackage{amsthm}
\usepackage[margin=1in, top=1.5in]{geometry}
\usepackage{booktabs}

\usepackage[slantedGreek]{mathptmx}
\usepackage{url}
\usepackage{bbm}
\usepackage{accents}
\newlength{\dhatheight}

\newtheorem*{theorem*}{Theorem}

\title[Adaptive market behavior]{
A model of adaptive, market behavior generating positive returns, volatility and system risk.}
\author{Misha Perepelitsa}

\thanks{Email: misha@math.uh.edu}

\date{\today}
\address{Misha Perepelitsa \\
misha@math.uh.edu\, 
University of Houston\\
PGH 631\\
4800 Calhoun Rd. \\
Houston, TX\\
USA}

\begin{document}

\begin{abstract}
We describe a simple model for speculative trading  based on adaptive behavior of economic agents.  
The adaptive behavior is expressed through a feedback mechanism for changing agents' stock-to-bond ratios, depending on the past performance of their  portfolios. The stock price is set according to the demand-supply for the asset derived from the agents'  target risk levels. 

Using the methodology of agent-based modeling we show that agents, acting endogenously and adaptively, create a persistent price bubble. The price dynamics generated by the trading process does not reveal any singularities, however the process is accompanied by growing aggregated risk that indicates increasing likelihood of a crash.

\end{abstract}

\maketitle

\begin{section}{Introduction}

We consider an agent-based model of economic behavior built on behavioral traits typical in the models of exchange:
a) every agent acts in his/her own interest;
b) agents have heterogeneous preferences;
c) agents interact and the interactions result in mutually optimal outcome;
d) agents' behavior is adaptive;
e) agents share a common belief that it is advantageous to participate in the market activity.
Traits a.--d. describe individual behavior, while the last condition, f., is a group property.

 The above conditions do not define a model in a unique way, as many different problems can fit into that framework, depending on the meaning  of  the terms ``preferences'', ``interactions'', ``optimal'', ``own interest'',  ``adaptive behavior'', and `` an outcome being advantageous''. For these terms we going to have precise definitions, originating in the analysis of speculative markets, that we describe in the paragraphs below. In short, the interactions will mean setting new price of an equity, agents' preferences will be defined by the stock-to-bond ratio, and adaptive behavior will be expressed through the  changes in  stock-to-bond ratios.

We will show that when the fundamental value of the traded asset is not readily estimated, being, for example, of speculative nature, the group of agents acting endogenously, according to conditions a.--f. can generate price dynamics with positive return. As there are no exogenous inputs in the model, the outcome  is clearly  a price bubble resulting from adaptive, self-centered behavior of agents. The phenomenon has some interesting features that make it different from the speculative bubbles due to actions of  ``chartists'' traders,  expected utility maximizers, or due to the  herding behavior of traders.  

For one thing, the price bubble is persistent, as all  agents in the group find it beneficial to carry on, provided they ignore mounting ``systematic risk'' of a breakdown (crash) of the market activity. We note here, that the systematic risk is a group property and it is typically not  observed by the agents, and thus not factored into the agents' short-term decision making process. On a long run the agents will eventually take the risk into consideration, and the price bubble will terminate. In section \ref{sec:discussion} we briefly discuss different scenarios of how that process may proceed.

Additionally, the model produces price fluctuations with some statistical similarity to fluctuations of real markets. A price chart  generated by the adaptive model is a graph growing, or decreasing, at a constant rate with some persistent but not increasing  volatility. Charts like that are not uncommon for the market indexes over a span of few years.   We conjecture that the adaptive type of behavior described in this paper contributes substantially to long term growth dynamics in speculative markets.

Models of adaptive economic behavior were developed by many authors in many different contexts. In a strategic decision making  they were advanced as  models of convergence by Cross \cite{Cross}, models of learning in games by Fundenberg-Levine\cite{FL}, reinforcement learning models by Roth-Erev \cite{RE}, to mention just a few contributions.  

The structure of agents' interactions in the model that we consider in this paper differs from the models in the above-mentioned references. Here,  agents use a singe strategy, or to be more precise {\it all} agents use  the {\it same} rule for the change of their strategies, and non-equilibrium price dynamics is generated by heterogeneity of agents' strategies caused by the updating rule. 
In our study we adopt the methodology of multi-agent dynamic simulations in the spirit of works of Stigler \cite{Stigler},  Kim-Markowitz \cite{KM},
Arthur-Holland-LeBaron-Palmer-Tayler \cite{AHLPT}, Levy-Levy-Solomon \cite{LLS4}.

%The adaptive strategy used by the agents is of the simply time: agents take into account only one, market, portfolio and act on the principle ``more is better''.  There are other ways that adaptation can be expressed. For example, agent can take into consideration not only  the market portfolio, but also the stocks that make it up. In section \ref{} we give an example of a market consisting two stocks, and show that  ``adaptive hedging'' behavior, built on the agents' belief that two stocks are negatively correlated, may result  in negative correlation in the returns of the traded  stocks.  

\subsection{Motivation}
The importance of the behavioral aspects of speculative markets has always been acknowledged. It is hardy deniable that 
irrationality of market participants and ``animal spirits'' are ubiquitous features that drive price instabilities and critical events. Attempts to understand the price changes in stock markets lead to the introduction of agent-based models that mimic the behavioral patterns of different groups of traders. Terms like ``fundamentalists'', ``noise-traders'',  ``portfolio re-balancers'', ``portfolio insurers'', ``adaptive learners'' have been introduced  into the models.
Kim-Markowitz \cite{KM} have shown the destabilizing effect of constant portfolio insurers on the price dynamics and proposed it as an explanation for the market crush of 1987.
The analysis of multi-agent models in Levy-Levy-Solomon \cite{LLS1, LLS2} shows that the rational traders (fundamentalists) do not out-evolve the irrational traders (chartists), and the later group will cause the price dynamics to go through a series of booms and crushes, that has statistical similarities with the empirical data. A number of non-linear differential equation models were proposed in Lux \cite{L1, L2, L3} and Lux-Marchesi \cite{L4, L5} to describe the evolution of the population of different types of traders, their interactions, the price dynamics  and changes in agents' wealth.  The model generates complex chaotic dynamics with the clustering of volatility of returns, fat tails in the distribution of  returns and wealth, as well as the long-term memory. Even if one restricts the traders to be rational, expected utility maximizers, the price will persistently develop  so called rational bubbles introduced in Mandelbrot \cite{M} and Blanchard-Watson\cite{BW}. Thus, the significant deviation from the fundamental value is the generic property of speculative prices.

This paper focuses on the behavior  of the market as a whole, consisting of many kinds stocks. The statistical properties of market portfolios are different from that of a single  stock. Looking at index charts, from after the Great Depression and onward, it is not unusual to see long periods of the scale 5-10 years when the market indexes grow at nearly constant rates, with mild, non-increasing volatility, and returns that fluctuate randomly around the mean. 

The positive returns, admittedly, are influenced by macro-economic parameters, such as the return on US government bonds. But the quantitative expression for the dependency or for the deviations is not clear.  We suggest that some of the features of "the happy growth" market dynamics can be generated endogenously by market participants following adaptive strategies.  Moreover, we suggest that  such behavior contributes to dramatics events that inevitably end the period of optimistic growth. To provide explanations we analyze an agent-based model in which interactions between agents mimic actual trading processes. The model is built on the following assumptions.

In the changing market environment traders constantly have to revise their investment portfolios, moving funds between "safer" and "riskier" assets, and to {\it change} their risk levels. We will assume that this re-balancing and updating of portfolios completely determines the dynamics of the market as a whole. To be more specific we postulate that  a) the price of the market portfolio is set by agents willing to re-balance funds between their stock and a safe, cash or bond, asset; b) when re-balancing, agents act solely on the basis of their stock-to-bond  ratios; c) agents change their investment ratios adaptively to the changes in their portfolios. 

The purpose of this paper is to understand the market dynamics described by the rules a.-c.,  isolating this type of behavior  from other market factors.  The main finding is a formula \eqref{BG}  on page 5 for the mean return in the excess of the safe investment return $r$ as a function of the systematic risk, that is, the risk of a crash, imposed by the collective behavior of the group.

\end{section}

\begin{figure}[htb]
\centering
  \begin{tabular}{@{}c@{}}
    \includegraphics[width=12cm]{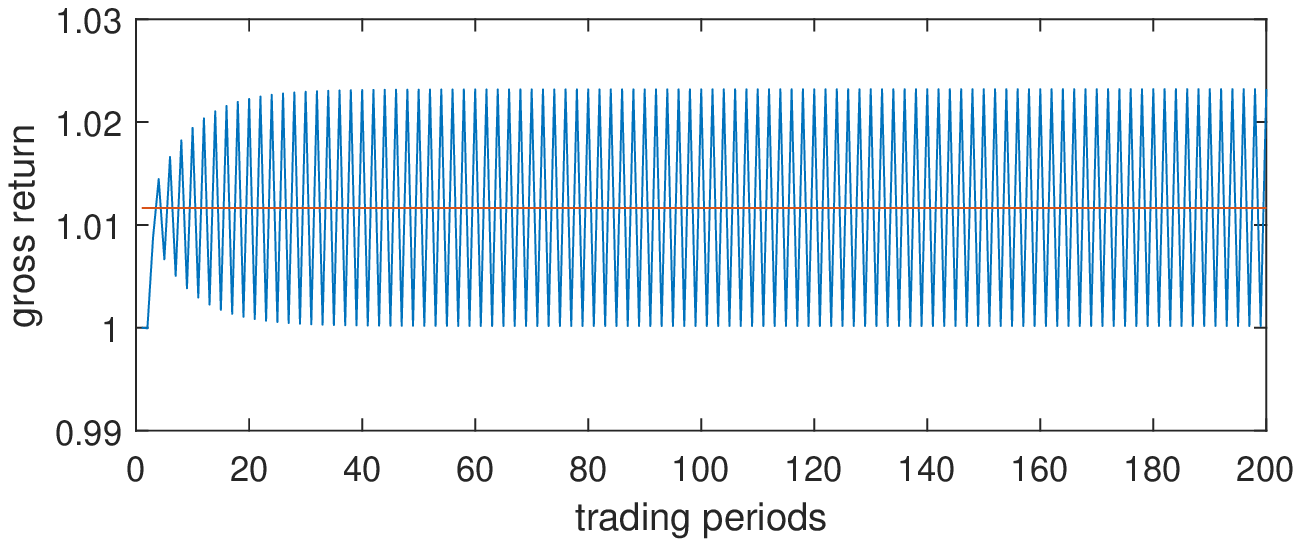} \\
    \includegraphics[width=12cm]{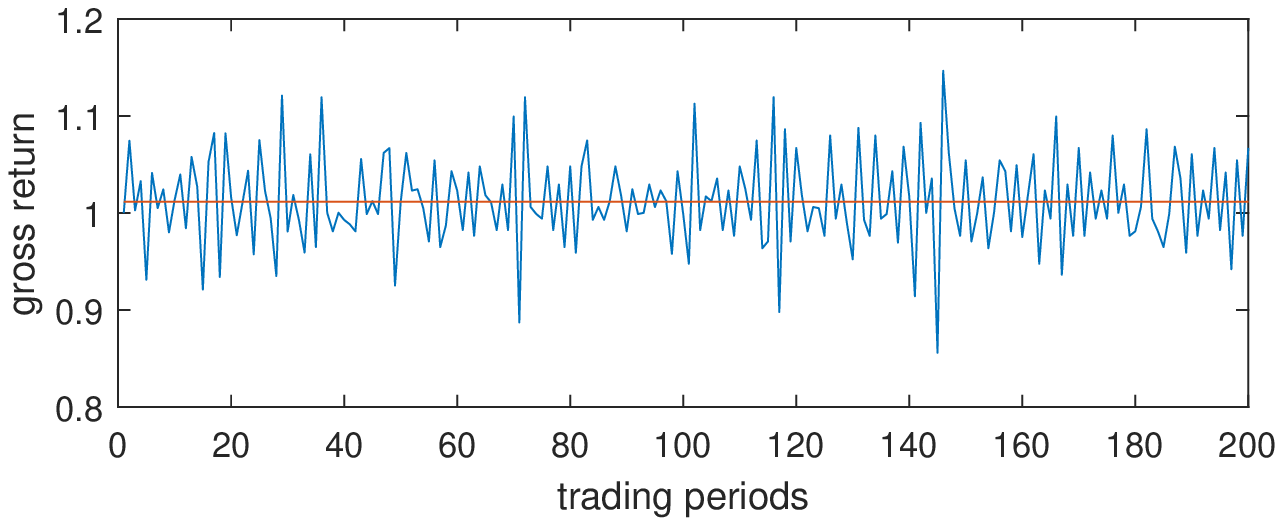} \\
    \includegraphics[width=12cm]{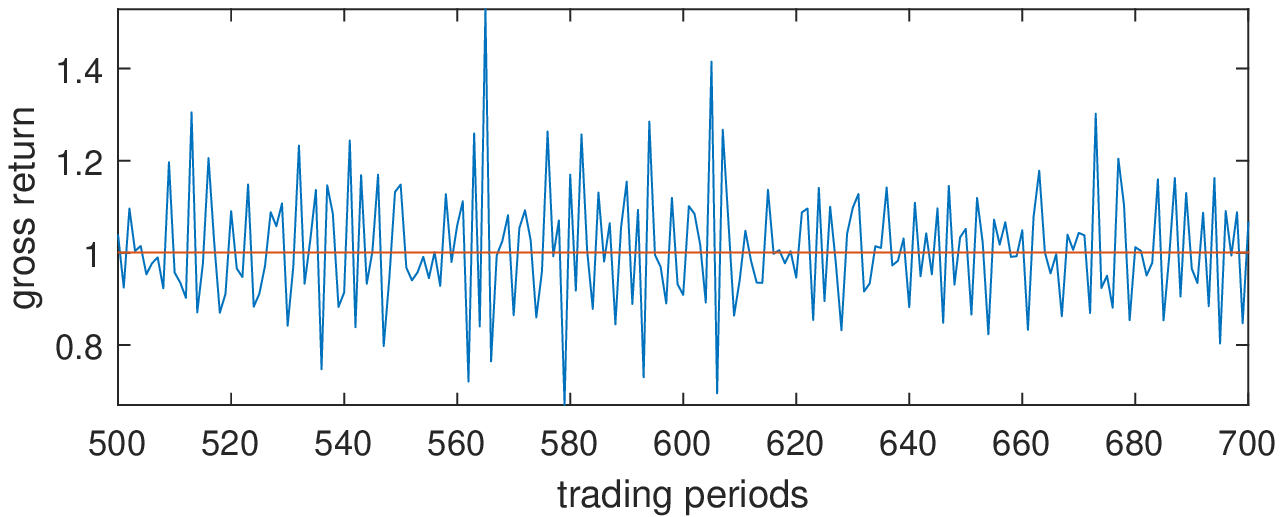} 
  \end{tabular}
  \caption{Gross returns during 200 trading periods. Top figure: deterministic 2-agent model.    Middle figure: 2-agent model with noise; graph shows only the returns from Petra and Paula trading sessions. Bottom figure: endogenous, multi-agent model with  $N=500$ agents and   $m=10$ active traders. The bottom chart shows 200 trading periods, after returns become stationary.  In all models  $\alpha=3.01,$ $\beta=0.34.$ Red lines are the geometric mean returns computed from \eqref{eq:return}:
1.012, 1.012 and 1.0006 for the top, middle and bottom chart, respectively. \label{fig:return}}

\end{figure}

\begin{figure}[htb]
\centering
\includegraphics[width=10cm]{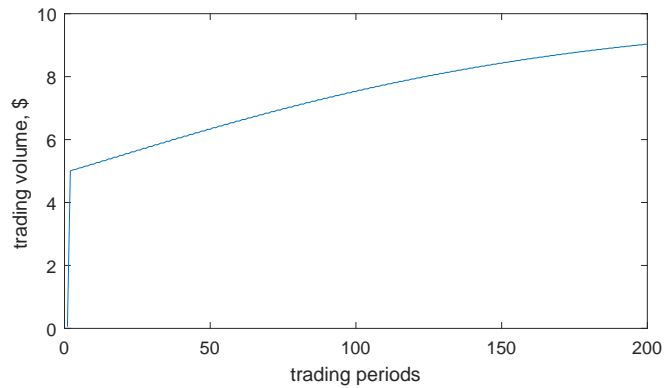} 
  \caption{Dollar value of shares traded per period. 
  \label{fig:Volume}}

\end{figure}

\begin{section}{The model}
To introduce the model we start with a market of only two agents trading shares of a single asset. The agents, call them Petra and Paula, do not know the asset fundamental value, but expect that the price can grow at moderate rates for long periods of time, and they always prefer increasing value to decreasing.

The agents meet at time intervals that we label with $n=0,1,2..$ At period $n,$ Petra's portfolio has   $(s^n_1,b^n_1),$  dollar value of stock and bond investments,
and Paula's: $(s^n_2,b^n_2).$ The stock price per share at period $n$ is $P^n.$ The next period investment portfolio $(s^{n+1}_i,b^{n+1}_i)$ and price $P^{n+1}$ determined trough the following steps.

Evaluating the last change in their portfolios, agents set target stock-to-bond ratios for the next period, that we denote by $k^{n+1}_i.$ The new price $P^{n+1}$ is set in such a way that the dollar amount of funds that Petra wishes to move from stocks to bonds, to be at the target ratio $k^{n+1}_1,$ equals the dollar amount Paula wants to move from bonds to stock to be at her ratio $k^{n+1}_2.$ This balance expressed as
\begin{equation}
\label{balance}
\frac{P^{n+1}}{P^n}s^n_1{}-{}k^{n+1}_1b^n_1{}={}k^{n+1}_2b^n_2{}-{}\frac{P^{n+1}}{P^n}s^n_2.
\end{equation}
Here we assuming that any fractional amount of a share can be exchanged, and for simplicity of the exposition, consider the case with zero interest in the bond account. With the newly set price the agents re-balance portfolios
% contain new amounts:
%\[
%s^{n+1}_1{}={}2\frac{P^{n+1}}{P^n}s^n_1{}-{}k^{n+1}_1b^n_1,\quad s^{n+1}_2{}={}2\frac{P^{n+1}}{P^n}s^n_2{}-{}k^{n+1}_2b^n_2,
%\]
%\[
%b^{n+1}_1{}={}\frac{P^{n+1}}{P^n}s^n_1{}+{}(1-k^{n+1}_1)b^n_1, \quad
%b^{n+1}_2{}={}\frac{P^{n+1}}{P^n}s^n_2{}+{}(1-k^{n+1}_2)b^n_2, \quad
%\]
and evaluate the market performance by comparing ratios $P^{n+1}s^{n}_i/(P^nb^n_i)$ to target values $k^{n+1}_i.$ If $P^{n+1}s^{n}_i/(P^nb^n_i)>k^{n+1}_i,$ from the agent $i$ point of view, the market performed better than expected, and if $P^{n+1}s^{n}_i/(P^nb^n_i)<k^{n+1}_i$ it performed worse.  Since when one agent is selling shares the other one is buying,  Petra and  Paula have different opinions on the market performance. The agents update the target portfolio ratios according to the rule
\begin{equation}
\label{eq:feedback}
k^{n+2}_i{}={}\left\{
\begin{array}{ll}
\alpha k^{n+1}_i & \dfrac{P^{n+1}s^{n}_i}{P^nb^n_i}>k^{n+1}_i,\\
\\
k^{n+1}_i & \dfrac{P^{n+1}s^{n}_i}{P^nb^n_i}=k^{n+1}_i,\\
\\
\beta k^{n+1}_i &\dfrac{P^{n+1}s^{n}_i}{P^nb^n_i}<k^{n+1}_i.
\end{array}
\right.
\end{equation}
With $\alpha>1,$ $0<\beta<1,$ we refer to this rule as {\it an adaptive feedback mechanism}. If the agent identifies a growing market she increases her stock-to-bond ratio by a fixed amount, while the other agent decreases her ratio. The feedback reflects the fact that when faced with a series of bad investments the agent will reduce the equity part of the portfolio, while with the investment growing better than expected, the agent will take riskier position. 

It easy to see that there is an equilibrium when no shares are traded and price doesn't change: $s^n_i=s^0_i,$ $b^n_i=b^0_i,$ $k^n_i=s^0_i/b^0_i,$ $P^n=P^0.$ Small deviations will set off a non-trivial dynamics. In this process, Petra and Paula will alternatively change their preferences by factors $\alpha$ and $\beta,$ so that, on average, the stock-to-bond  ratios change at the rate $(\alpha\beta)^{1/2}$ per period. The gross return, $P^{n+1}/P^n,$ after a short transient period will settle at the average value of $(\alpha\beta)^{1/2}$ per period, but will keep oscillating around that value,
see the top graph in Figure \ref{fig:return}. The stock price goes up when agents have bias toward the increase of the risk preferences, expressing the optimistic outlook, quantified by the condition $\alpha \beta>1.$ 

The story of this process is simple: Petra and Paula are trading stock and
and move their stock-to-bond ratios up and down with an upward average trend. This  drives up the average demand and consequently the price. 

Figure \ref{fig:Volume} shows a simulated, traded dollar value of stock per period. The volume keeps growing up to a certain limit -- the amount of cash available to agents. At the same time, as the stock price increases exponentially, the number of traded shares will approach zero.

To make a model more viable we can arrange the trading mechanism so that agents do not need to interact directly or even know about each other. They might be matched by a third party, a market maker, to whom they submit their order books. Moreover, the stock price might change in between the periods when Petra and Paula are trading, due to, for example, actions of other agents. The middle graph in Figure \ref{fig:return} shows a realization of the stock price when  it is changing as a geometric random walk between Petra and Paula trading sessions. The outcome is still the same: Petra and Paula are pushing price up, The rate of increase, per unit of time, depends now on how many times during that periods they trade.   

To make the model completely endogenous we populated  the market by many copies of Petra and Paula. At every trading period there will be a group of "active" traders willing to re-balance their portfolios. They submit the book orders to the market maker who executes the orders and sets the new price, according to the rule similar to \eqref{balance}, see section \ref{Formulas} for details. Each active agent  uses rule \eqref{eq:feedback}  to update her stock-to-bond ratio taking into account only the performance of her portfolio. The process proceeds to the next step, with a new, randomly chosen group of active traders.
The model is determined by the following parameters: $N$--number of agents, $m$--number of active agents that is fixed or can be chosen from a distribution, $\tau$--the number of trading periods per year, the feedback mechanism \eqref{eq:feedback} and the set of initial portfolios together with the starting price $P_0$ per share. With such parameters, each agent, on average, re-balances his portfolio $m\tau/M$ times per year, which we set to be of order 1.

A typical realization of price dynamics is shown in Figure \ref{fig:log_price}. Statistically, the returns quickly move to a stationary regime with no significant long term correlations. The bottom graph in figures \ref{fig:return} shows the realizations of the gross return.   The product $\alpha\beta$ still determines if the stock price is pumped up or down. The formula for the mean return per year
\begin{equation}
\label{eq:return}
E[R]/r=\left(\alpha \beta\right)^{\frac{m\tau}{2N}},
\end{equation}
where $r$ is the return on cash account, appears to be a good quantitative approximation for the geometric mean of return on the stock investment, see Perepelitsa-Timofeev \cite{PT} for a technical report on numerical simulations of the model.
\end{section}

\begin{figure}[htbp]
   \centering
       \includegraphics[width=10cm]{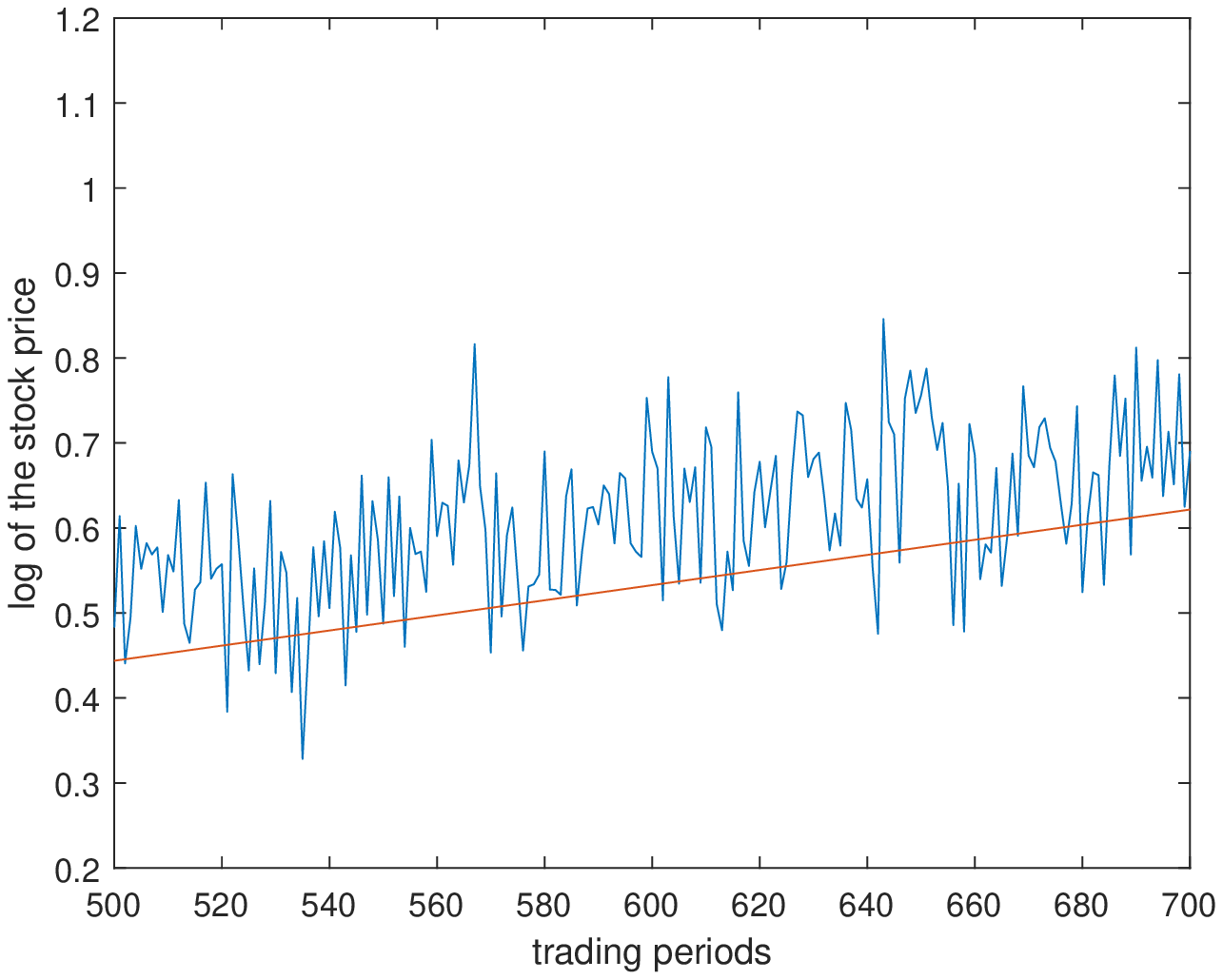}
   \caption{Logarithm of the stock price, after the returns become stationary. Straight line is the mean return line of slope \eqref{eq:return}. }
   \label{fig:log_price}
\end{figure}

\begin{section}{Discussion}
\label{sec:discussion}
Besides the price charts, agent-based models provide detailed information on the wealth distribution among agents. This information allows one to study the effects of market activity on the population of agents. As an example, let us consider trading between $N=500$ agents with the parameters used in simulation presented in Figure \ref{fig:log_price}. The stock price is going up at the rate $14\%$ per year with the standard deviation of $0.08.$ In 10 years, agents that started with the identical portfolios and stock-to-bond ratios, have their portfolios distributed according to Figure \ref{fig:scatter}.

\begin{figure}[htbp]
   \centering
       \includegraphics[width=10cm]{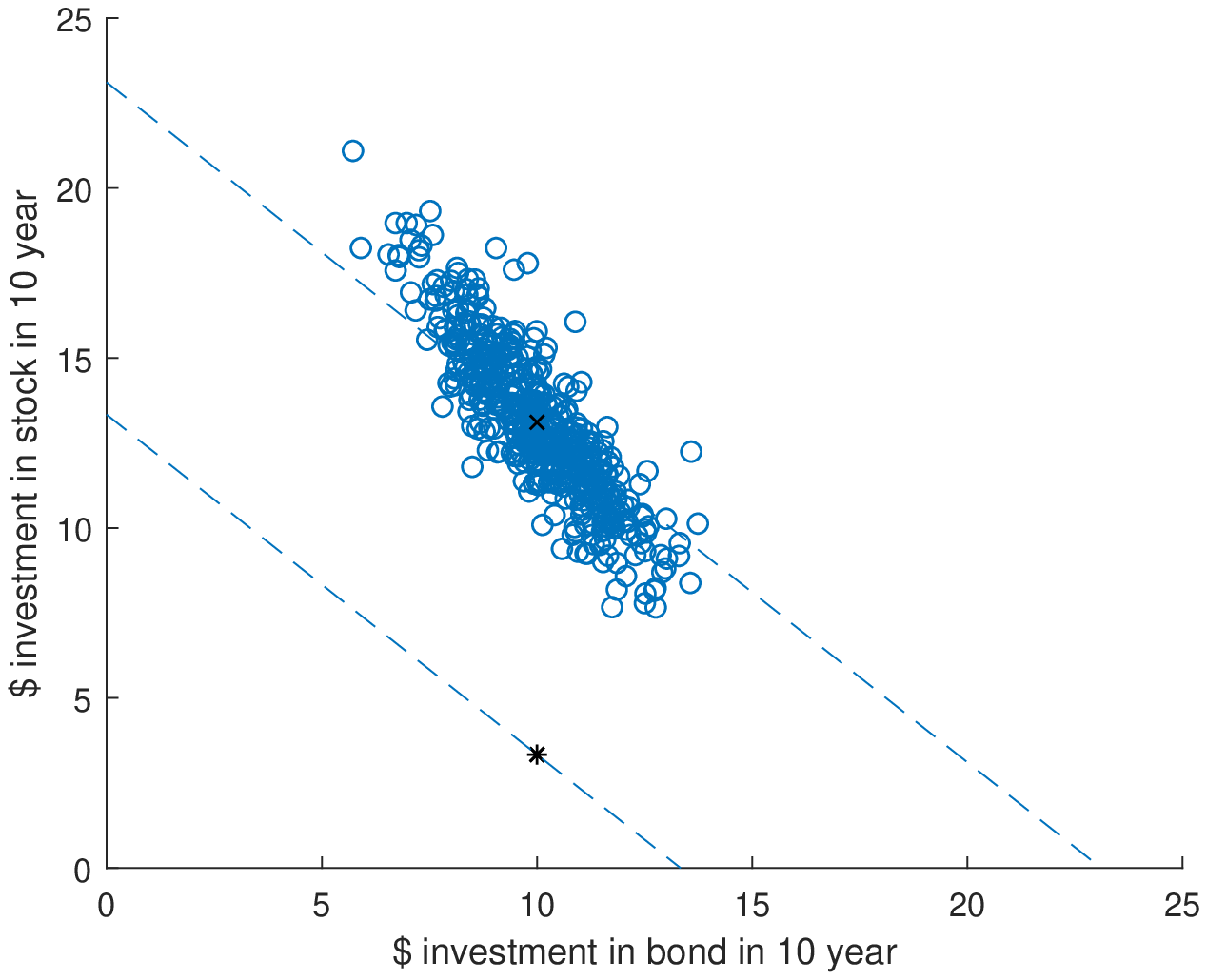}
   \caption{Scatter plot of agents' portfolios in 10  years. Lines of slope -1 are the lines of the constant total wealth. The initial portfolios of all agents is marked with '*'.  The mean portfolio is marked with 'x'. }
   \label{fig:scatter}
\end{figure}

Note that in total money (stock+bond) {\it all} agents are better off participating in the market than holding everything in cash. Moreover, almost a half of them have increased their bond accounts as well as stock investments. This however comes at the price of all agents adopting more riskier positions, as measured by their stock-to-bond ratios. After 10 years, the mean ratio in the population is 5, compared to its initial value of 1/3. The minimum of the ratio over the group is 2. The group becomes significantly riskier. We can relate this risk to the systematic risk of the market breakdown. Indeed, every trading period, a real-life  agent would  face a decision to continue trading using the strategy \eqref{eq:feedback} or start selling the stock, in hope to recover its cash value. The outcome would depend on what other agents are doing. It can be expressed by a version of a Prisoner's Dilemma, Table \ref{table:game}, in which the agent plays against the rest of the group and chooses between ``staying'', continuing using adaptive strategy,   or ``selling," selling of all stock shares. As the  payoffs in the game grow the agent will eventually prefer to use his dominant strategy ``sell.'' With the rest of the agents following the same path, the market would crash.

\begin{table}
\centering
\begin{tabular}{cc}
 & Market \\
Agent &
\begin{tabular}{@{}lll@{}}
\toprule
   &  stay     & sell  \\
stay  & $b_0+0.1s_0E[R]^n$   & $0$   \\
sell   & $b_0+s_0E[R|^n$ & $b_0$  \\
\bottomrule
\end{tabular}

\end{tabular}

\vspace{15pt}

\caption{Schematic decision matrix for Agent vs. Market game with the payoffs to the agent in $n$ years : $b_0$ is value of the agent's bond account.  $s_0E[R]^n$ is the stock account of the agent. The agent recovers all of it in sell-stay scenario. The payoff  stay-stay is uncertain; the agent expecting to recover  cash value for at least a fraction of her stock investment. 
 \label{table:game}}

\end{table}

The annual growth rate of the mean of the stock-to-bond ratios can be estimated in terms of the parameters of the model as
$
\left(\alpha \beta\right)^{\frac{m\tau}{2N}}.
$ With this we can interpret  \eqref{eq:return} as
\begin{equation}
\label{BG}
E[R]/r=\mbox{the rate of increase of  the risk of a market breakdown};
\end{equation}
the risk premium of the market portfolio is proportional to the rate of increase of the systematic risk. Interestingly, a similar formula, relating the return to the increasing  risk of a crash, applies to  a rational bubble, described in Blanchard-Watson \cite{BW}. 

As an alternative scenario to a crash, the group may switch to a different  feedback mechanism \eqref{eq:feedback}  with  a "pessimistic"  bias, determined by the condition  $\alpha\beta<1.$ The bubble will deflate gradually to lower price levels. It can be deflated all the way to the starting price $P_0$ and the only result of  this trading cycle will  be the redistribution of initial funds among agents.

Gradual deflation is certainly a better alternative to a crash, as the former is always accompanied by a panic behavior that has adverse effects on the economy.  Interestingly, from the group prospective,
both a crash and a deflation trading might have similar end results.
Figure \ref{fig:scatter_crash} shows simulated distributions of portfolios after a crash and a 2-year deflation trading.  The scatter plots of both distributions are clearly comparable.  
However, to deflate the bubble  with the adaptive feedback \eqref{eq:feedback} requires agents, from time to time, to buy shares when the stock  price is clearly decreasing. This condition might go counter with  primarily individualistic preferences of agents, leading to a crash.

\begin{figure}[htbp]
   \centering
       \includegraphics[width=10cm]{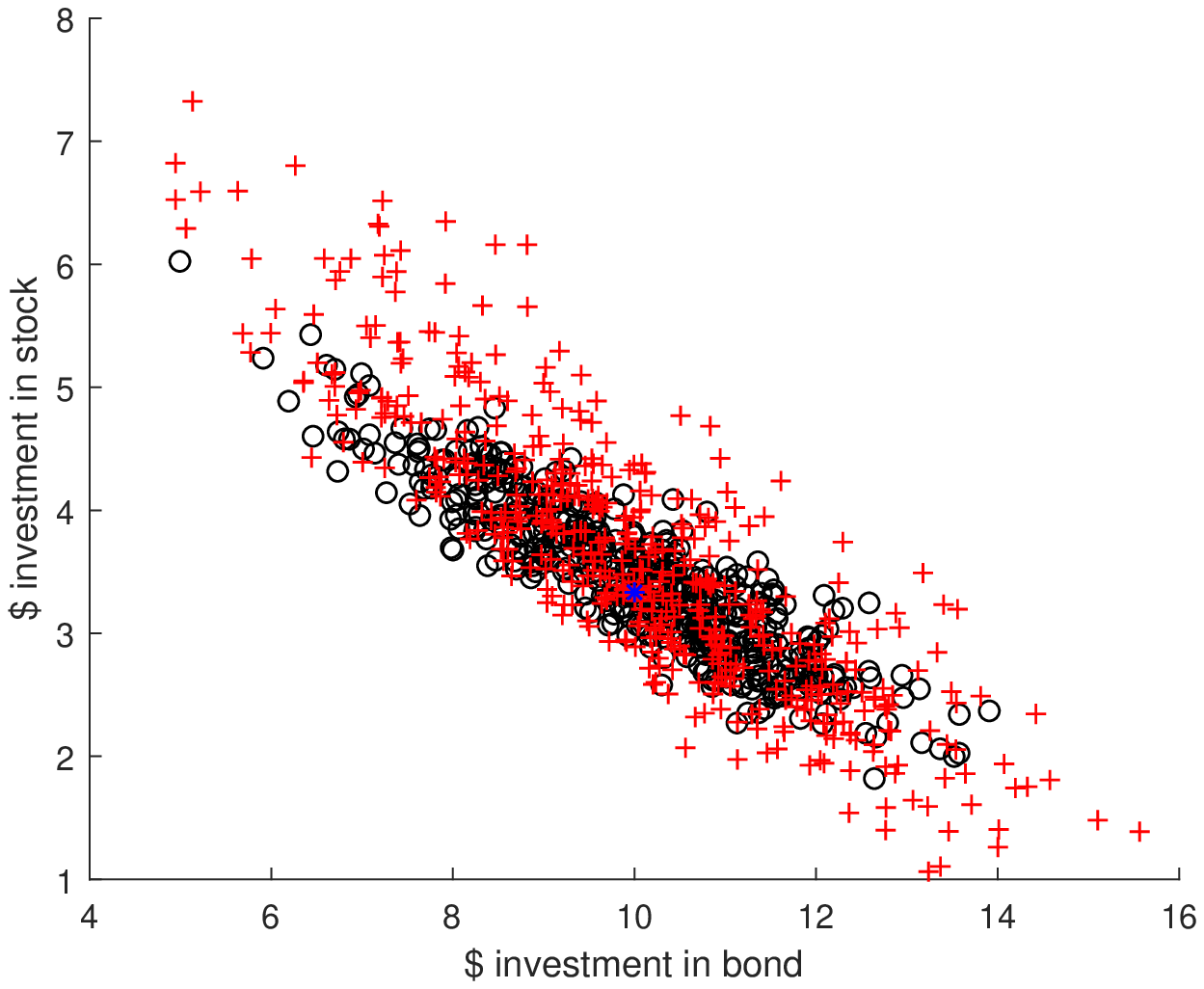}
   \caption{Scatter plot of agents' portfolios in 10  years of a growing bubble followed by a crash or a gradual deflation: `o' represent portfolios after a crash to the starting level of stock price $P_0$; `+' represent portfolios after 2 years of deflating bubble dynamics that brings the price down to $P_0;$ `*' -- portfolios of agents at time $t=0.$ }
   \label{fig:scatter_crash}
\end{figure}

%The second observation from Figure \ref{fig:scatter} concerns the spread of the risk ratios. It can be seen from the scatter plot that nearly half of the group, not only increases the total wealth, but also ends up with more money in cash account than it started with. Since the cash is conserved, the other half of the group will have less. The population is clearly heterogeneous. When the price is deflated the agents from high risk-low cash group will sell stock and agents with higher cash 

Let  us finally mention that when the bubble is considered as a part of a bigger market environment there is also a possibility for it turning into a kind of a self-aggregated Ponzi scheme. After all, a long-term stable growth of the bubble will attract new participants. While the influx of new money keeps growing at  the same rate as the return generated by the bubble,  the  agents can realize the profit by selling stocks without dropping the price. The price bubble will prolong its existence but its fate will be determined by the balance of in- and out-flux of money.

\end{section}

\begin{section}{Appendix: Properties of the model}
\label{Formulas}
Consider a set of $N$ agents, described by their portfolios $(s_i,b_i),$ $i=1\ldots N,$ of dollar values of a stock and cash accounts, and let $k_i$ stand for agent $i$ stock-to-bond ratio, either actual, or a future, target value. $P_0$ will denote a current price per share and $P$ the new price determined by agents' demand. Let $\{i_l\,:\,l=1\ldots m\}$  be the set of  ``active'' agents, i.e. the ones setting the new price. The set of active traders is determined each trading period by a random draw from the population.

If $x_{i_l}$ is the dollar amount that agent $i_l$ wants to invest in stock, then
\[
\frac{\frac{P}{P_0}s_{i_l}+x_{i_l}}{b_{i_l}-x_{i_l}}{}={}k_{i_l}.
\] 
 The demand-supply balance is
\[
\sum_{l=1}^m x_{i_l}{}={}0,
\]
which can be solved for $P:$
\[
\frac{P}{P_0}{}={}\left( \sum_{l=1}^m\frac{k_{i_l}b_{i_l}}{1+k_{i_l}}\right)\left(\sum_{l=1}^m\frac{s_{i_l}}{1+k_{i_l}}\right)^{-1}.
\]
During the trading, the total amount of all cash accounts is conserved, as well as the number of shares of stocks owned by agents.  Note here, that we are assuming that any fractional amount a share can be traded. Once the price is set, agents move corresponding amounts between cash and stock accounts, re-balancing their portfolios. The update mechanism for new stock-to-bond ratios is given by \eqref{eq:feedback}.  The interaction is repeated the following trading periods with randomly selected sets of active agents. As in 2-agent model, there is an equilibrium solution, when all agents have balanced portfolios, $s_i/b_i=k_i,$ $i=1,\ldots,N,$ and stock price doesn't grow, $P_{n}=P_{n-1}.$
If money in saving accounts grow at the rate $r,$ the equilibrium price will grow at the same rate $P_n=rP_{n-1}.$ When the initial data are out of the equilibrium, the system will exhibit non-trivial dynamics, diverging from the equilibrium.

In the non-equilibrium regime, the stock price increases or decreases according, together with the mean stock-to-bond ratios. The return, however, becomes nearly {\it stationary}, with the geometric mean approximately equal to \eqref{eq:return}, see \cite{PT}. The distribution of returns depends on the mechanism for selecting a random group of active agents. If the number of active agents, $m,$ is fixed the returns in the stationary regime follow log-normal distribution. If number $m$ is itself random, for example chosen from an uniform distribution, the returns have significantly more mass at the tails and at 1, than a corresponding log-normal distribution. Moreover, the returns show no significant correlations over the lags of two or more trading periods, while the next period return is negatively correlated with the present one, see \cite{PT} for details.

Of interest are also statistical properties of the price dynamics for a collection of processes with different values of parameters $(\alpha,\beta).$
 Table \ref{table:correlation} presents the correlation between mean returns and standard deviations. For each pair of values we compute the stationary mean positive or negative return and the standard deviation, and consider the distribution of these values for a large number of different values $\alpha,\beta$ in the vicinity of $1.$ The numbers do depend on other parameter of the model, such as $m$ and $N,$ but they indicate positive correlation between mean positive return and the standard deviation, and negative correlation between mean negative return and standard deviations. This property is observed in the returns of real markets.

\begin{table}
\centering
\begin{tabular}{@{}lll@{}}
\toprule
	       &  positive  returns     &  negative returns\\

\midrule

m=5          &  0.1 &            \\
m=10        &  0.07 & -0.09   \\
m=50		 &  0.21  & -0.47    \\
S\&P500  	 &   0.22 & -0.54    \\
\bottomrule
\end{tabular}

\vspace{15pt}

\caption{Correlation between mean positive (negative) returns and standard deviations. The adaptive model with  $N=500$ agents and $m=5, 10, 50$ active traders was used to compute  the mean returns and standard deviations for 1000 different values of $\alpha,\beta\in[0.8, 1.5].$ For S\&P500, 85 pairs of the mean daily returns and standard deviations computed from non-overlapping, semi-annual periods of the index from the period 1960-2010.  
 \label{table:correlation}}

\end{table}

% We will proceed heuristically to derive the formula for the mean return of the stock. Following formula \eqref{rate:stock} we assume that the rate of return $r_s$ equals the product of rate of return on bond, $r,$ and geometric mean of the change of stock-to-bond ratio of agents per trading period. The probability that an agent is selected as ``active'' is the fraction $\frac{m}{N}.$ Assuming that his/her changes in the riskiness, $\alpha,$ or $\beta$ are equally likely, the geometric mean equals $(\alpha\beta)^{\frac{m}{2N}},$ and we arrive at the formula
%\begin{equation}
%\label{rate:stock_N}
%r_s{}={}r(\alpha\beta)^{\frac{m}{2N}}.
%\end{equation}
%\vspace{7pt}

%\subsection{Trading on margin} We'd like to describe here a variation of the trading model that accounts for the market behavior known as trading on margins. The behavior was wide spread prior to. Agents borrow money to buy stock using the current value of stock as a collateral. 

\end{section}

\end{document}